\documentclass
[pra,aps,twocolumn]{revtex4}
\usepackage{amssymb}
\usepackage{latexsym}

\def\be{\begin{equation}}
\def\ee{\end{equation}}
\def\bea{\begin{eqnarray}}
\def\eea{\end{eqnarray}}
\def\Tr{{\rm \,Tr\,}}

\def\Tr{{\rm \,Tr\,}}

\def\r{{\bf r}}
\def\p{{\bf p}}

\def\sgn{\,{\rm sgn\,}}

\def\h2m{\frac{\hbar^2}{2m}}
\def\p0{{P_{\beta H^0_N}}}

\begin{document}

\title{{\flushleft{\small {\rm Published as
Phys. Rev. Lett. {\bf 94}, 080402 (2005)}\\}}
\vspace{1cm}
\large\bf Equivalence of Bose-Einstein condensation
and symmetry breaking}
\author{Andr\'as S\"ut\H o\\
Research Institute for Solid State Physics and Optics\\ Hungarian Academy of
Sciences \\ P. O. Box 49, H-1525 Budapest\\ Hungary\\
E-mail: suto@szfki.hu}
\thispagestyle{empty}
\begin{abstract}
\noindent
Based on a classic paper by Ginibre [Commun. Math. Phys. {\bf 8} 26 (1968)]
it is shown that whenever Bogoliubov's approximation, that is, the replacement
of $a_0$ and $a_0^*$ by complex numbers in the Hamiltonian, asymptotically yields
the right pressure, it also implies the asymptotic equality of
$|\langle a_0\rangle|^2/V$ and $\langle a_0^*a_0\rangle/V$ in symmetry breaking
fields, irrespective of the existence or absence of Bose-Einstein condensation.
Because the former was proved by Ginibre to hold for absolutely integrable
superstable pair interactions, the latter is equally valid in this case. Apart from
Ginibre's work, our proof uses only a simple convexity inequality due to Griffiths.

\vspace{2mm}

\vspace{2mm}
\noindent

\end{abstract}
\maketitle
The equivalence of Bose-Einstein condensation (BEC) and a spontaneous breakdown of the
gauge symmetry related to the choice of a global phase factor in creation and
annihilation operators is an intriguing
problem which apparently has not yet been solved rigorously.
Most of the time the equivalence
is tacitly assumed; for example, Hohenberg's celebrated result \cite{Ho} about
one- and two-dimensional Bose-systems is often considered as the proof of the absence
of BEC in low dimensions although it proves the absence of symmetry breaking.
The equivalence of BEC and symmetry breaking can be summarized as the asymptotic
equality of $|\langle a_0\rangle|^2/V$ and $\langle a_0^*a_0\rangle/V$ in the
limit of infinite volume, when the
thermal or ground state averages are taken in the presence of a symmetry breaking
field. The operator $a_0^*$ creates a boson in a one-particle state distinguished
by the fact that its occupation may become macroscopic.
The problem is that in finite volumes the first quantity is always less than the
second; if equality installs asymptotically, it is the work of a large deviation
principle resulting in the equivalence of certain statistical ensembles.
This question, closely related to the validity of the Bogoliubov approximation (BA)
 \cite{BogA},
raised only a limited interest almost exclusively among mathematical physicists
\cite{Gin}-\cite{ZB}.
The equivalence of the two notions in the above sense was recently shown by the
present author for some models with a simplified interaction \cite{Su},
without using BA. In that
paper it was overlooked that the equivalence for a general interacting Bose gas
was a more or less direct
consequence of a powerful result about BA
by Ginibre, obtained almost forty
years ago \cite{Gin}. One may say that Ginibre himself overlooked this important
implication of his work.
This note is to present the necessary argument.

In his classic but not widely known paper Ginibre considered a Bose gas with a
rather general pair
interaction making nearly the minimum assumption to guarantee a non-pathological
thermodynamic behaviour. The pair interaction $\phi$ is superstable and weakly
tempered, i.e. for $n$ particles in a box of volume $V$ the total potential
energy
\be
U(\r_1,\ldots,\r_n)=\sum_{i<j}\phi(\r_i-\r_j)\geq -bn+an^2/V
\ee
for any set of position vectors in the box, and
\be
\phi(\r)\leq \phi_0 r^{-(d+\epsilon)}
\ee
for $|\r|=r\geq R$. Here $a>0$ and $b$ are independent of the volume,
$d$ is the space dimension and
$\phi_0$, $\epsilon$ and $R$ are positive constants.
Ginibre was interested in proving the correctness of
BA which consists in treating $a_0$ and $a_0^*$ as complex numbers. Here
$a_0^*$ creates a boson in the one-particle state $\varphi_0\equiv 1/\sqrt{V}$
(but $\varphi_0$ could be any other one-particle state). Ginibre's interpretation
of BA is as follows. Let $F_0$ be the single-mode Fock space spanned by the product
states $\varphi_0^{\otimes n}$, $n=0,1,2,\ldots$ and let $F'$
be the Fock space built on the orthogonal complement of $\varphi_0$ in the one-particle
Hilbert space
\footnote{This is a corrected definition of $F'$, that agrees with Ginibre's original definition.
The correction is to be done also in the PRL.}.
Let, moreover, $P_{F'}$ be the orthogonal projection onto $F'$.
Then, for an operator $B$ on $F$ and a
complex number $C$ the Bogoliubov approximation of $B$ is
\be\label{BA}
B_0(C)=P_{F'}A_CBA_C^*P_{F'}
\ee
where it is understood that $B_0(C)$ acts only on elements of $F'$.
The operator
\be
A^*_C=e^{Ca_0^*-\overline{C}a_0}=e^{-|C|^2/2}e^{Ca_0^*}e^{-\overline{C}a_0}
\ee
applied to a $\psi$ in $F'$ creates the coherent state
\be
|C\rangle=e^{-|C|^2/2}\sum_{n\geq 0}\frac{C^n}{\sqrt{n!}}\,\varphi_0^{\otimes n}\ ,
\ee
tensor-multiplying $\psi$.
If $\psi_1,\psi_2$ are in $F'$,
\be
\langle\psi_1|B_0(C)|\psi_2\rangle=\langle\psi_1\otimes C|B|\psi_2\otimes C\rangle.
\ee
In the first part of his paper
Ginibre applied the transformation (\ref{BA}) onto the density matrix
$W=e^{-\beta H}$. The Hamiltonian of the system is
\be\label{H}
H=T+U-\mu N-\nu\sqrt{V}(a_0+a_0^*)\
\ee
where $T$ is the kinetic energy, $\mu$ is real, $N$ is the particle number operator
and the amplitude $\nu$ of the gauge symmetry breaking field is
chosen to be real. Because $A_C^*$ is unitary, without the projection the transformation
preserves norm, trace and positivity. Together with the projection the trace
of a positive operator cannot but decrease. So
\be\label{Wineq}
\Tr W_0(C)=\Tr'A_CWA_C^*\leq \Tr W=Z
\ee
where $\Tr'$ is the trace in $F'$ and thus
\be\label{pineq}
V^{-1}\log\Tr W_0(C)\leq V^{-1}\log Z \equiv\beta p_V(\mu,\nu)\ .
\ee
Ginibre proved that the limit of the pressure $p_V$ exists and
\be\label{W}
\lim_{V\to\infty}\sup_C\, (\beta V)^{-1}\log\Tr W_0(C)
=\lim_{V\to\infty} p_V\equiv p(\mu,\nu)
\ee
showing thereby that with the right choice of $C$, BA for the density matrix
reproduces the exact pressure in the thermodynamic limit.

Bogoliubov's original idea was to make the replacement in the Hamiltonian.
Because $a_0|C\rangle=C|C\rangle$, writing $H$
as a normal-ordered polynomial of creation and annihilation operators and applying
(\ref{BA}), the resulting $H_0(C)$ is indeed the same as the outcome of a simple
substitution.
In coherent states particles can enter in contact with each other.
While (\ref{W}) holds
also for hard-core interactions, to make $H_0(C)$ meaningful, hard cores have to
be excluded. With the additional condition of the absolute integrability of $\phi$,
Ginibre proved the analog of Eq.~(\ref{W}). The inequalities
\be
Z_0(C)=\Tr'e^{-\beta H_0(C)}\leq \Tr W_0(C)\leq Z
\ee
or
\be
p_0(C)=(\beta V)^{-1}\log Z_0(C)\leq p_V\ ,
\ee
derived in \cite{Gin},
suggest that, again, $p_0(C)$ is to be maximized. The main result (Theorem 3)
of \cite{Gin} is
\be\label{p0}
\lim_{V\to\infty}\sup_C\, p_0(C,\mu,\nu)=\lim_{V\to\infty} p_V(\mu,\nu)=p(\mu,\nu)
\ee
which is our point of departure.

The proof of the equivalence of BEC and symmetry
breaking in the sense discussed in the introduction is as follows. \\
(i)
When a sequence of convex functions $f_n$ converges (necessarily to a convex
function $f$), the
sequence $f_n'$ of derivatives also converges apart from possibly a set a zero
measure. In particular, the inequalities
\be\label{Grif}
f'(x-0)\leq \liminf_{n\to\infty}f_n'(x-0)\leq\limsup_{n\to\infty}f_n'(x+0)
\leq f'(x+0)
\ee
due to Griffiths \cite{Gr} hold true. There is at most a countable infinite
set of $x$ such that $f'(x-0)<f'(x+0)$; otherwise we have equalities in (\ref{Grif}).
In the first two applications below the subscript $n$ will correspond to $V$.\\
(ii)
We embed $H$ into a one-parameter family of auxiliary Hamiltonians
\be
H'=T+U-\mu N'-\mu_0 N_0-\nu\sqrt{V}(a_0+a_0^*)
\ee
where $N_0=a_0^*a_0$ and $N'=N-N_0$.
It is easily checked that the pressure $p_V'(\mu,\mu_0,\nu)$
corresponding to $H'$
has all the nice properties shown in \cite{Gin} for $p_V(\mu,\nu)$. Especially,
for fixed $\mu$ and $\nu$, $p_V'$ is a convex increasing and real analytic
function of $\mu_0$ (so it is continuous, $p_V'(\mu,\mu,\nu)=p_V(\mu,\nu)$),
and for fixed $\mu$ and $\mu_0$
it is a convex even and real analytic function of $\nu$.
These
properties, apart from the analyticity but including continuity,
are inherited by the (existing)  thermodynamic limit
$p'(\mu,\mu_0,\nu)$, so $p'(\mu,\mu,\nu)=p(\mu,\nu)$. Also,
\be
\frac{\langle a_0\rangle_{\mu,\mu_0,\nu}}{\sqrt{V}}
=\frac{\langle a_0^*\rangle_{\mu,\mu_0,\nu}}{\sqrt{V}}
=\frac{1}{2}\frac{\partial p_V'(\mu,\mu_0,\nu)}{\partial\nu}
\ee
and
\be
\frac{\langle N_0\rangle_{\mu,\mu_0,\nu}}{V}=\frac{\partial p_V'(\mu,\mu_0,\nu)}
{\partial\mu_0}
\ee
where the averages are taken with the density matrix
$e^{-\beta H'}/\Tr e^{-\beta H'}$.
For any fixed $\mu$ and for $(\mu_0,\nu\neq 0)$ in a set $\Omega_\mu$ of full
Lebesgue measure
both $\partial p'/\partial\nu$ and $\partial p'/\partial\mu_0$ exist, and by
Eq.~(\ref{Grif})
\be\label{anul1}
\lim_{V\to\infty}\langle a_0\rangle_{\mu,\mu_0,\nu}/\sqrt{V}
=(1/2)\partial p'(\mu,\mu_0,\nu)/\partial\nu
\ee
and
\be\label{nnul1}
\lim_{V\to\infty}\langle N_0\rangle_{\mu,\mu_0,\nu}/V
=\partial p'(\mu,\mu_0,\nu)/\partial\mu_0\ .
\ee
With the possible exception of a countable number of values of $\nu$
Eq.~(\ref{anul1}) holds also for $\mu_0=\mu$
.
This may not be true for
Eq.~(\ref{nnul1}): there is an abstract possibility that for a positive-measure
set of $\nu$ the $\mu_0$-derivative of $p'$ does not exist at $\mu_0=\mu$.
This,
however, would have
no practical importance because for {\em any} choice of $(\mu,\nu)$,
$\partial p'/\partial\mu_0$ exists for almost every $\mu_0$ and, hence, for $\mu_0$
arbitrarily close to $\mu$. We return to this point later.\\
(iii)
Apply BA to $H'$ to obtain $H_0'(C)$, $Z_0'(C)$ and $p_0'(C)$. Theorem 3 of \cite{Gin}
extends without any further ado to this case, resulting $p_0'(C)\leq p_V'$ and
\be\label{p0prim}
\lim_{V\to\infty}\sup_C\, p_0'(C)=p'
\ee
for all $\mu$, $\mu_0$ and $\nu$. With the choice
$\nu\neq 0$ real the maximum is attained for $C$ real, $C\nu\geq 0$. Within this
set $\cal{C}$, $p_0'$ reads
\be
p_0'(C,\mu,\mu_0,\nu)=p_0'(C,\mu,0,0)+\mu_0\frac{C^2}{V}+2|\nu|\frac{|C|}{\sqrt{V}}\ .
\ee
For fixed $C$ and $\mu$ this is a convex function of both $\mu_0$ and $\nu$;
\be
\sup_{C}p_0'(C)=p_0'(C_{\rm max}(\mu,\mu_0,\nu),\mu,\mu_0,\nu)
\ee
is the upper envelope of
$\{p_0'(C,\mu,\mu_0,\nu)|C\in\cal{C}\}$ and is, therefore, convex in $\mu_0$ and
$\nu$. The finite-volume pressure $p_0'(C,\mu,0,0)=p_0'(-C,\mu,0,0)$ is a real
analytic function of $C$. So $p_0'(C,\mu,0,0)\approx p_0'(0,\mu,0,0)+aC^2$ close to
$C=0$. Thus, if $\nu\neq 0$, $C_{\rm max}\neq 0$ either,
and $\partial p_0'/\partial C=0$ at $C=C_{\rm max}$. This implies
\be\label{partnu}
\frac{\partial p_0'(C_{\rm max}(\mu,\mu_0,\nu),\mu,\mu_0,\nu)}{\partial\nu}
=2\sgn\nu\,\frac{|C_{\rm max}|}{\sqrt{V}}\phantom{aaaaaaaa}
\ee
and
\be
\frac{\partial p_0'(C_{\rm max}(\mu,\mu_0,\nu),\mu,\mu_0,\nu)}{\partial\mu_0}
=\frac{C_{\rm max}^2}{V}\ .
\ee
Now we use (\ref{Grif}) a second time, with $f_n=p_0'(C_{\rm max},\mu,\mu_0,\nu)$
and $f=p'(\mu,\mu_0,\nu)$. 
For any $\mu$ and any $(\mu_0,\nu)\in\Omega_\mu$
\be\label{cmax}
c'_{\rm max}(\mu,\mu_0,\nu)=\lim_{V\to\infty}|C_{\rm max}(\mu,\mu_0,\nu)|/\sqrt{V}
\ee
exists, and
\be\label{equiv1}
\lim_{V\to\infty}\frac{\langle a_0\rangle_{\mu,\mu_0,\nu}}{\sqrt{V}}
=\sgn\nu\,c'_{\rm max}={\rm sgn\,}\nu\,
\lim_{V\to\infty}\sqrt{\frac{\langle N_0\rangle_{\mu,\mu_0,\nu}}{V}}\ .
\ee

\vspace{1mm}
\noindent
(iv)
The above result can be improved.
When $\partial p(\mu,\nu)/\partial\nu$ exists, we can set $\mu_0=\mu$ in
Eqs.~(\ref{anul1}), (\ref{partnu}) and (\ref{cmax}) and obtain the first half of
Eq.~(\ref{equiv1}),
\be\label{dp/dnu}
\frac{1}{2}\frac{\partial p(\mu,\nu)}{\partial\nu}=
\lim_{V\to\infty}\frac{\langle a_0\rangle_{\mu,\nu}}{\sqrt{V}}
=\sgn\nu\,c_{\rm max}(\mu,\nu)\ .
\ee
Here $c_{\rm max}(\mu,\nu)=c'_{\rm max}(\mu,\mu,\nu)$
which is clear from $p_0(C,\mu,\nu)=p_0'(C,\mu,\mu,\nu)$.
Now fixing any $\mu$, for all $\nu$ outside a possible ($\mu$-dependent)
set of zero measure there exist both
the derivative $\partial p(\mu,\nu)/\partial\nu$ and a sequence $\mu_n\downarrow\mu$
such that $(\mu_n,\nu)\in\Omega_\mu$. For such a $\nu$ and $\mu_n$,
$p'(\mu,\mu_n,\nu)\to p(\mu,\nu)$ by continuity and
$\partial p'(\mu,\mu_n,\nu)/\partial\nu\to\partial p(\mu,\nu)/\partial\nu$
by Griffiths' Lemma (\ref{Grif}).
Combining Eqs.~(\ref{anul1}), (\ref{equiv1}) and (\ref{dp/dnu}) one obtains
that for almost all $(\mu,\nu\neq 0)$
\bea\label{equiv2}
\lim_{V\to\infty}\frac{\langle a_0\rangle_{\mu,\nu}}{\sqrt{V}}
=\sgn\nu\,c_{\rm max}(\mu,\nu)\phantom{aaaaaaaaaa}\nonumber\\
={\rm sgn\,}\nu\,\lim_{n\to\infty}
\lim_{V\to\infty}\sqrt{\frac{\langle N_0\rangle_{\mu,\mu_n,\nu}}{V}}\ .
\eea

Equation (\ref{equiv2}) is nearly the final result except that in the right
member the two limits should be interchangeable. This, however, is easily seen.
Because of the Schwarz inequality,
\be
\lim_{V\to\infty}\frac{|\langle a_0\rangle_{\mu,\nu}|}{\sqrt{V}}
\leq\lim_{V\to\infty}\sqrt{\frac{\langle N_0\rangle_{\mu,\nu}}{V}}\ .
\ee
Suppose that there is a strict inequality here. Note that
$\langle N_0\rangle_{\mu,\mu_0,\nu}$ is an increasing function of $\mu_0$.
Then for some $\delta>0$ and for all $V>V_\delta$ and all $\mu_0\geq\mu$
\be
\frac{|\langle a_0\rangle_{\mu,\nu}|}{\sqrt{V}}+\delta
<\sqrt{\frac{\langle N_0\rangle_{\mu,\nu}}{V}}\leq
\sqrt{\frac{\langle N_0\rangle_{\mu,\mu_0,\nu}}{V}} \ .
\ee
The contradiction is obvious because $\mu_n\downarrow\mu$ in Eq.~(\ref{equiv2}).
Our final result is, thus, that for almost all $(\mu,\nu\neq 0)$
\be\label{equiv3}
\lim_{V\to\infty}\frac{\langle a_0\rangle_{\mu,\nu}}{\sqrt{V}}
=\sgn\nu\,c_{\rm max}(\mu,\nu) 
={\rm sgn\,}\nu\,
\lim_{V\to\infty}\sqrt{\frac{\langle N_0\rangle_{\mu,\nu}}{V}}\ .
\ee
Equation (\ref{equiv3}) proves that an optimized c-number substitution provides
the right answer at least for the asymptotic value of $\langle a_0\rangle/\sqrt{V}$
and $\langle a_0^*a_0\rangle/V$. It does not tell anything about these values
(although for $\nu\neq 0$ they are surely of order 1, as in \cite{Su}),
especially, about their limit when $\nu$ tends to zero. So the equation
does not prove that there is a spontaneous symmetry breaking, but it does prove
that symmetry breaking occurs simultaneously with Bose-Einstein condensation.

In approximate calculations one often starts with the canonical transformation
$a_0=b_0+\langle a_0\rangle$ where, by definition, the
operator $b_0$ has zero mean; see e.g. \cite{SzK}.
The approximations made afterwards are based on the hypothesis that the fluctuations
of $b_0$ are negligible or small compared with $\langle a_0^*a_0\rangle$.
Equation (\ref{equiv3}) justifies this hypothesis by proving
$\langle b_0^*b_0\rangle_{\mu,\nu}/V\to 0$ as $V\to\infty$.

The functional form of $p'$ is
\bea
p'(\mu,\mu_0,\nu)=g(c_{\rm max}(\mu,\mu_0,\nu),\mu)\phantom{aaaaaaaaaaaaaa}
\nonumber\\
+\mu_0 c_{\rm max}^2(\mu,\mu_0,\nu)+2|\nu| c_{\rm max}(\mu,\mu_0,\nu)\ .
\eea
Alternately, $p'$ has to satisfy the equation
\be\label{diffeq}
\frac{\partial p'(\mu,\mu_0,\nu)}
{\partial\mu_0}=
\frac{1}{4}\left(\frac{\partial p'(\mu,\mu_0,\nu)}{\partial\nu}\right)^2
\ee
obtained by reading the two ends of Eq.~(\ref{equiv1}).
Apart from the fact that $\partial p'/\partial\nu$ tends to $\partial p/\partial\nu$,
as $\mu_0$ tends to $\mu$, the Bose gas described by
the extended Hamiltonian $H'$ may have its own interest. In any case,
if $\mu_0$ is close to $\mu$, the behaviour of this system is not expected to
significantly differ from that of the original one.
BEC or a spontaneous symmetry breaking is described by solutions of
(\ref{diffeq}) that are
convex in all the three variables and have a discontinuous $\nu$-derivative
at $\nu=0$.

A different solution to the problem treated in this Letter can be found in
Ref. [12].

This work was partially supported by OTKA Grants T 042914 and T 046129.

\end{document}